\documentstyle[manuscript,aps]{revtex} 

\begin{document}
 
\newcommand{\NP}[1]{Nucl. \ Phys.  {\bf #1}}
\newcommand{\CMP}[1]{Commun. \ Math. \ Phys.  {\bf #1}}
\newcommand{\JMP}[1]{J. \ Math. \ Phys.  {\bf #1}}
\newcommand{\PL}[1]{\ Phys.  \ Lett.  {\bf #1}}
\renewcommand{\thesection}{\arabic{section}}
\begin{titlepage} 
\begin{center}
\preprint{hep-th/9605099}
\title{Chiral Quantization of the WZW $SU(n)$ Model}
\author{L. Caneschi}
\address{Dept. of Physics, Univ. of Ferrara and INFN Ferrara, Italy}
\author{M. Lysiansky \footnote{E-mail: famisha@wishful.weizmann.ac.il}}
 
\address{Dept. of Physics of Complex Systems, Weizmann Institute of Science,
Rehovot, Israel}
\maketitle 
\end{center}
\begin{abstract}
 
We quantize the $SU(n)$ Wess-Zumino-Witten model in terms of left and right
 chiral variables choosing an appropriate gauge and we compare our results
 with the results that have been previously obtained in the algebraic
treatment of the problem. The algebra of the chiral vertex 
operators in the fundamental representation is verified by solving  
appropriate Knizhnik-Zamolodchikov equations. 
\end{abstract}
\end{titlepage}
\section{Introduction}\label{intr}

 Rational conformal field theories are well understood from the
 algebraic point of view \cite{BPZ}\cite{TK}\cite{MS}. 
  The structure is determined by the existence of
   commuting chiral algebras ${\cal A}\times {\bar{\cal A}}$ . The Hilbert
space consists of the
direct product of infinite dimensional representations of these algebras,
 while the chiral vertex   operators
  are intertwining operators between these
representations. The way in which the left and right representations are
 paired is determined by modular invariance . The chiral operators
 are coupled together to produce "physical operators" which
  are local with respect to each other. 
 
    It    is an interesting  question
    to study in which way the aforementioned
 structure is encoded into the lagrangian describing the conformal
 theory, when such a description exists. 
      This problem was studied for the Wess-Zumino-Witten (WZW)
  class of conformal theories where the chiral algebra is the
  Kac-Moody one. These theories are described by the
 WZW  lagrangian \cite{W} . The basic idea \cite{Blo}\cite{Fad}\cite{Sch} in
 establishing the connection between the lagrangian and
 the algebraic descriptions is to quantize the theory  in a
 parameterization of the phase space which is chiral, i. e. the traditional
 field and conjugate momentum are replaced by group valued functions
 representing the left and right moving classical solutions. 
 
   This approach was further developed in \cite {G1}. In the present
 work we continue the study of this problem for the SU (n) WZW theory. 
  The main results obtained are~:
 
   a) A complete description of the Hilbert space required by the
  chiral quantization which is shown to be identical to the one required
  by the algebraic treatment. 
 
  b)The exact correspondence between the chiral vertex operators
  of the algebraic treatment and the chiral group elements of
  the lagrangian one. 
 
   c)The correspondence between the mutually local physical operators
  and appropriate combinations of chiral group elements. 
 
    In solving the problems mentioned above    we found very useful
  the chiral quantization proposed in \cite{Blo}\cite{Fad}\cite{G1} of a "toy
model "  ,
  the motion on a group manifold. 
 
    In Section 2 we solve completely in a chiral quantization
  the motion on a SU (n) group manifold. A technical problem which appears in
the chiral quantization is the appropriate choice of a gauge. We compare the
chiral solution and the conventional one and we  prove explicitly that our
choice leads to the correct mapping between chiral and standard operators. 
   Since the zero modes of the WZW model are described
  by the chiral lagrangian of the "toy model" this indicates
   that an analogous gauge choice is justified also for the WZW model. 
 
    In Section 3  we rederive the lagrangian representing the WZW
   model in chiral variables. We give an explicit description of the
    Hilbert space and of the matrix elements of the chiral group
    elements and using the Knizhnik-Zamolodchikov equations
    we prove that the chiral group elements satisfy the correct
    commutation relations. 
 
      In Section 4 we summarize the results and list the problems
  that are still  unsolved.

\section{Chiral quantization of the motion  on a group manifold}
\subsection{The standard treatment}\label{2A}
 
Let us consider the free
motion   on the compact simple Lie group G. We define
$t^{a}=\{H^{i}, E^{\alpha} \}$ for $i=1, . . . , r$, where $r$ is the rank of G and
$\alpha$ are its  roots to be a basis of the generators for the
corresponding  Lie
 algebra  $\cal G$ . We  choose the normalization such that $tr (t^{a}
 t^{b})=\delta_{a b}$. The action for the motion
 of a "free particle" on the group manifold is~:
\begin{eqnarray}
S=-{1 \over 2} \int tr (g^{-1} {\dot g} g^{-1} {\dot g}) d t, \hskip 1cm g \in G
\label{act1}
\end{eqnarray}
This action is invariant under
transformations belonging to the group  $G \times G$,
corresponding to the multiplication of $g$ with arbitrary group
elements from the left  and right respectively. 
 
  The Hilbert space is spanned by the group elements. Alternatively
 \cite{M}  one can use the dual   basis  of matrix elements
 of all the unitary irreducible
 representations labeled by the highest weight
 of the  representation $\lambda$ and two set  of "magnetic"
 quantum numbers which we denote collectively by  $\mu_l$ and
 $\mu_r$ respectively. 
 
   The hamiltonian is diagonal in the dual basis ,
   the eigenvalue  $E_{\lambda}$
 being the second order Casimir of the group which depends just
 on the highest weight~:
 \begin{eqnarray}
   E_{\lambda}=\lambda (\lambda +\varrho)
 \end{eqnarray}
 where $\varrho$ is the half sum of the positive roots. 
  Since the hamiltonian is diagonalized it is easy to calculate the
 propagator   $K (g (t), g (0))$
 from an initial state  characterized by an eigenvalue
 of $g$ , $g (0)$ to a final state $g (t)$~:
\begin{equation}
   K (g (t), g (0))= \sum_{\lambda , \mu_l ,\mu_r} D_{\mu_l , \mu_r}^{\lambda *}
 (g (t)) D_{\mu_l , \mu_r}^{\lambda} (g (0)) exp (i t E_{\lambda})
\label{prop}
\end{equation}
   In the next subsection we  quantize this simple model in the chiral
  quantization and compare the results with the above formulae.

\subsection{Classical analysis in chiral variables}\label{2B}
 
 Let us consider the equations of motion corresponding to the action
 (\ref{act1})~:
 
\begin{equation}
{d \over {d t}} (g^{-1} {\dot g})= {d \over {d t}} ({\dot g} g^{-1} ) =0
\end{equation}
We can write the solution in the form~:
 
\begin{equation}
g (t)=u e^{i {\vec p} {\vec H} t} v, \hskip 1cm u, v \in G
\end{equation}
There is an ambiguity in the parametrization since replacing
\begin{eqnarray}
u\rightarrow uh, \hskip 1cm v\rightarrow h^{-1}v
\label{ambig}
\end{eqnarray}
 
where $h$ is an arbitrary element from the maximal
torus $T$ leads to the same solution. 
 
The equations of motion can also be  written in the Hamiltonian form~:
 
\begin{equation}
\omega =  g^{-1} {\dot g}, \hskip 0.5 cm  \omega \in {\cal G}
\end{equation}
\begin{eqnarray*}
 {\dot \omega} =0
\end{eqnarray*}
These equations can be obtained from the action~:
\begin{equation}
S=- \int tr (\omega g^{-1} {\dot g}-{\omega^{2} \over 2}) d t
\label{frorac}
\end{equation}
The action (\ref{frorac}) is  first
 order  and  $\omega $ , $g$ are the phase-space variables. 
 
To factorize the left and right theories we make the
 change of variable from $\omega$ and $g$ to $u$ and $v$. The phase space
is the space of all classical solutions. In the standard treatment $g$ ,
$\omega$ describe this space by representing the initial conditions on a
classical trajectory $g=g (t)$~:
 
\begin{equation}
\omega= g^{-1} {\dot g} |_{t=0}
\label{2.2}
\end{equation}
 
\begin{eqnarray*}
g=g (t)|_{t=0}
\end{eqnarray*}
 Using  the chiral parametrization of the solution (\ref{2.2}) and (\ref{2.2})
we obtain~:
 
\begin{equation}
\omega=i v^{-1} {\vec p}{\vec H} v
\label{2.3}
\end{equation}
\begin{eqnarray*}
g=u v
\end{eqnarray*}
Here $ {\vec p} $ is a  periodic variable , the periodicity being defined
by the root lattice. In the new variables the action is~:
 
\begin{equation}
S=-i \int tr ({\vec p}{\vec H} (u^{-1} {\dot u}+ {\dot v}v^{-1}-i {{{\vec p}
{\vec H}} \over 2})) d t
\label{actbgf}
\end{equation}
The ambiguity in the parametrization mentioned above (\ref{ambig}) manifests
itself
as a gauge invariance of the action (\ref{actbgf}) under the transformation
 (\ref{ambig}). As a consequence the symplectic $\Omega$ form derived from the
action (\ref{actbgf})~:
\begin{equation}
\Omega=-i tr (d {\vec p}{\vec H} (u^{-1} {d u}+d v v^{-1})-{\vec p}{\vec H}
 (u^{-1} {d u} u^{-1} {d u}+d v v^{-1} d v v^{-1})
\end{equation}
 is degenerate  and cannot be inverted. 
 
To avoid this problem we fix the  gauge in a convenient way
introducing  new variables
 $ \vec p_{l}$ and $ \vec p_{r}$  in terms of which the action takes the form~:
 
\begin{equation}
S= -i \int  tr ({\vec H} \vec p_{l} u^{-1}\dot{u}+{\vec H} \vec p_{r} \dot{v}
v^{-1} -{i \over 8} ( (\vec p_{l}+\vec p_{r}){\vec H})^2 ) d t
\label{plr}
\end{equation}
The relevant symplectic  form now is non-degenerate~:
 
\begin{equation}
\tilde{\Omega} =-i tr (d \vec p_{l} \vec H u^{-1}du-\vec p_{l}
\vec H u^{-1}uu^{-1}u
 +d \vec p_{r} \vec H dvv^{-1}+\vec p_{r} \vec H dvv^{-1}dvv^{-1})
\end{equation}
The action (\ref{plr}) is  equivalent to (\ref{actbgf}) provided that the
constraint~:
\begin{equation}
\vec \phi =\vec p_{l} -\vec p_{r}=0
\label{constr}
\end{equation}
 is fulfilled. Equation (\ref{constr}) defines a set of abelian first class
constraints and therefore its Poisson brackets (or commutators in the quantum
 case ) with physical
quantities vanish. In the quantum case these constraints are imposed with an
appropriate projection on the physical Hilbert space. 
 
The simplest proof of the equivalence between (\ref{plr}) supplemented with
 (\ref{constr}) and (\ref{actbgf})
is obtained introducing
the $\delta-$function representing the constraint (\ref{constr}) in the path
 integral with the action (\ref{actbgf}). By explicit solution of the
model we will prove the equivalence in the operatorial formulation.

We can obtain the Poisson brackets by inverting $\tilde{\Omega}$. The result
is~:
\begin{mathletters}
\label{2.5}
\begin{equation}
\{ u_1 u_2 \}= u_{1} u_{2} R_{12}^{l}, \hskip 0.8 cm R_{12}^{l}=
\sum_{\alpha \in \Phi} {i \over {p_{l} \alpha}} E_{\alpha} \otimes
 E_{- \alpha}
\end{equation}
\begin{eqnarray}
\{ v_1 v_2 \}= v_{1} v_{2} R_{12}^{r}, \hskip 0.8 cm  R_{12}^{r}=
\sum_{\alpha \in \Phi} {i \over {p_{r} \alpha}} E_{- \alpha} \otimes
 E_{\alpha}
\end{eqnarray}
\begin{eqnarray}
 \{ \vec {p}_l, u \}=-i u \vec H, \hskip 0.8 cm  \{ \vec {p}_r, v \}=
-i \vec H v
\end{eqnarray}
\end{mathletters}
The Poisson brackets between left and right variables vanish. 
The left (right) symmetry generators are given by~:
\begin{mathletters}
\begin{equation}
J^a_l =itr (t^a u\vec{p_l} \vec{H}u^{-1})\hskip 0.3 cm  , \hskip 0.3 cm
\{ J^a_l , J^b_l \}=f^{abc}J^c_l
\end{equation}
\begin{eqnarray}
J^a_r =itr (t^a v^{-1}\vec{p_r} \vec{H}v)\hskip 0.3 cm  , \hskip 0.3 cm
\{ J^a_r , J^b_r \}=f^{abc}J^c_r
\end{eqnarray}
\end{mathletters}
and the transformations properties of $u, v$ under the group action are~:
\begin{equation}
\{ J^a_l , u \}=it^a u \hskip 0.2 cm  , \hskip 0.2 cm  \{ J^a_r , v \}=ivt^a
\end{equation}

\subsection{Quantization}\label{s22}

Quantization of the previous Poisson brackets (\ref{2.5})
 gives the following result \cite{Fad1}\cite{G2}~:
 
\def\br#1#2{#1_{1} #1_{2}=#1_{2} #1_{1} B_{12}^{#2}}
\begin{mathletters}
\label{2.6}
\begin{equation}
u_1 u_2 =u_2 u_1 B^l_{12}, \hskip 0.8 cm  v_1 v_2 =B^r_{12}v_2 v_1
\end{equation}
\begin{eqnarray}
[ p^{i}_{l}, u]= \hbar u H^{i}, \hskip 0.8 cm  [ p^{i}_{r}, v ]= \hbar H^{i} v
\end{eqnarray}
\begin{eqnarray}
[ p^{i}_{l}, p^{j}_{l}]=[ p^{i}_{r}, p^{j}_{r}]=0
\end{eqnarray}
\end{mathletters}
The explicit solution for the B-matrices was obtained in\cite{G2} for the case
 when $u$ , $v$ are matrices in the
 fundamental representation of $G=SU (n)$. Let $\Phi$ denote the space of roots
 and $\vec \lambda_{i}$ for $i=1.. . n$ be the weights in the fundamental
 representation, with highest weight  $\vec \lambda_{1}$. The normalization
 of these vectors is chosen to be $\vec \lambda_{i} \vec \lambda_{j}=
\delta_{ij}-{1 \over n} $. The  Cartan-Weyl basis is given by~:
 
\begin{equation}
\vec H=\sum_{i} \vec \lambda_{i} e_{ii}
\end{equation}
and the step operators
$E_{\alpha_{ij}}$ are represented by the matrices $ (e_{ij})_{ab}=\delta_{ia}
\delta_{jb}$ for $\alpha_{ij}=\vec \lambda_{i}-\vec \lambda_{j}
 \in \Phi$ . In this basis the matrices $B_{l} (B_{r})$
 have the following form~:

\begin{mathletters}
\label{2.7}
\begin{equation}
B_{l,r}=exp (-\sum_{\alpha} \theta_{\alpha}^{l,r} E_{\alpha} \otimes
 E_{-\alpha} ) =\sum_{\alpha} (\cos \theta_{\alpha}^{l,r}E_{\alpha} E_{-\alpha}
 \otimes E_{-\alpha}E_{\alpha}-\sin \theta_{\alpha}^{l,r} E_{\alpha} \otimes
 E_{-\alpha} )
\end  {equation}
 where $  \sin
\theta_{\alpha}^{l}={\hbar \over {p_{l} \alpha}}$ and
 $\sin \theta_{\alpha}^{r}=-{\hbar \over
{p_{r} \alpha}}$
 
\end{mathletters}
The symmetry generators can be derived from these commutators. Taking into
account the corrections due to the ordering problem they have the form~:
 
\begin{eqnarray}
J_{l}^{a}=itr ( t^{a} (u (p_{l} H) u^{-1}+2 \hbar H^2)), \hskip 0.8cm
J_{r}^{a}=i  tr (t^{a} (v^{-1} (p_{r} H) v+2 \hbar H^2))
\label{gener}
\end{eqnarray}
 
The generators (\ref{gener}) fulfil the standard commutation relations~:
\begin{eqnarray}
[J_{l}^{a}, J_{l}^{b}]=f^{abc} J_{l}^{c},\hskip 0.8cm [J_{r}^{a},J_{r}^{b}]=
f^{abc} J_{r}^{c}
\end{eqnarray}
 
The transformation properties of the $u$ and $v$ operators are defined by~:
\begin{mathletters}
\label{uj}
\begin{equation}
 [J_{l}^{a}, u]=i \hbar t^{a} u
\end{equation}
\begin{eqnarray}
 [J_{r}^{a}, v]=i \hbar v t^{a}
\end{eqnarray}
\end{mathletters}
The second-order  Casimir operators for left and right theories depend
 only on  $p_{l}^{i}$ and $p_{r}^{i}$ respectively~:
\begin{equation}
C_{l, r}=p^{2}_{l, r} -{\hbar}^2 tr (H^2 H^2)
\end{equation}
We  can now construct the
 Hilbert space for these theories. We will consider only the left theory,
 because the right one is completely analogous~:
 
\begin{equation}
{\cal H}=\bigoplus_{\{ \lambda\}}{\cal H}_{\lambda}
\end{equation}
where ${\cal{H}}_{\lambda}$ is
 the irreducible unitary representation with highest weight ${\lambda}$. 
 To describe these spaces it is convenient
 to introduce the following notations. Every irrep  of $SU (n)$ is
 characterized by a partition~:
 
\begin{equation}
[m]=[m_{i n }]=[m_{1 n}, m_{2 n}, . . . , m_{{n-1}, n}, 0]\label{2.8}
\end{equation}
of $n-1$ non-negative integers, obeying the relation~:
 
\begin{equation}
m_{i n} \ge m_{i+1, n}
\end{equation}
An elegant notational convention to label in a unique way a
 state of the irrep $[m_{in}]$ is the Gelfand pattern\cite{J}, denoted by $ (m)$. 
It consists of a triangular
 array  of $\hskip 0.2 cm {{n (n+1)} \over 2}-1 \hskip 0.2 cm$
integers $\{ m_{ij} \}$, $1\leq i\leq j \leq n$ with $m_{nn}=0$ which satisfy
the "betweenness conditions"~:
 
\begin{equation}
m_{i, j+1} \ge m_{i, j} \ge m_{i+1, j}
\label{betw}
\end{equation}
The k-row,
 $ (k \not =n)$, of this pattern is a highest weight for some irrep of
 $U (k)$. The possibility of such a description of states is based on the 
Weyl branching law for $U (j)$. It asserts that in the restriction from group
to subgroup $U (j+1)\rightarrow U (j)$ for any given irrep of $U (j+1)$ with
highest weight $[m_{i,j+1}]$ the irrep of $U (j)$ with highest weight
$[m_{ij}]$ satisfying the "betweenness condition" (\ref{betw}) occurs only
once . Therefore the Gelfand pattern corresponds to the branching chain
of restrictions from group to subgroup~:
\begin{eqnarray}
SU (n)\rightarrow U (n-1)\rightarrow \cdots \rightarrow U (1)
\label{chain}
\end{eqnarray}

We define now the action of the operators in the chiral theory on the Hilbert
space. The operators $p^{i}$ commute with all the generators $J^{a}$ . Therefore
they depend only on the highest weight. Their explicit action on the state is~:
 
\begin{equation}
p^{i}| (m) \rangle=\hbar (m_{i n} +n -i) | (m) \rangle
\end{equation}
 
We construct now the action of the $u_{ij}$ operators . 
From equation (\ref{uj}) we see that the $u_{ij}$'s are tensor
operators in the fundamental representation and the $i$ index designates the
magnetic quantum number. 
 
From the commutation relations~:
\begin{equation}
[p^i, u]=\hbar uH^i
\end{equation}
it follows  that
\begin{equation}
\langle (\tilde{m})|u_{ij}| (m)\rangle\sim\prod_{k=1}^{n-1}\delta_
{\tilde{m}_{kn}, m_{kn}+\delta_{kj}-\delta_{kn}}
\label{rind}
\end{equation}
Therefore the  meaning of the right index of $u_{ij}$
is completely different from the meaning of the left one~:  $u_{ij}$ carries an
arbitrary vector belonging to irrep $[m]_n$ into irrep $[\tilde{m}]_n$~:
\begin{eqnarray}
\tilde{m}_{kn}=m_{kn}+\delta_{kj}-\delta_{jn}
\end{eqnarray}
We will relate the $u_{ij}$ to the unit tensor operators of $SU (n)$\cite{B}. 
The unit tensor operator is labeled by two Gelfand patterns, denoted by~:
\begin{equation}
\left\langle\begin{array}{ccc}
 (\Gamma)_{n-1}\\\
[M]_{n}\\
 (M)_{n-1}
\end{array}\right\rangle
\label{utp}
\end{equation}
In this notation the lower pattern
\begin{equation}
 (M)_n\equiv\left (\begin{array}{rr}
[M]_n\\
 (M)_{n-1}\end{array}\right)
\end{equation}
identifies  the transformation properties in $SU (n)$ of the tensor
operator. The upper pattern~:
\begin{equation}
 (\Gamma)_n\equiv\left (
\begin{array}{rr} (\Gamma)_{n-1}\\\
 (M)_{n}\end{array}\right)
\end{equation}
 is of the same form as a Gelfand pattern and satisfies the same "betweenness
conditions"~:
\begin{equation}
\Gamma_{i, j+1}\geq\Gamma_{ij}\geq\Gamma_{i+1, j+1}
\end{equation}
$ (\Gamma)_n$ designates the fact that the tensor operator carries an
 arbitrary vector, belonging to irrep $[m]_n$ into a vector, belonging
 to irrep $[\tilde{m}]_n$ of $SU (n)$, where~:
\begin{equation}
\tilde{m}_{in}=m_{in}+\Delta_{in} (\Gamma)
\label{shift}
\end{equation}
\begin{eqnarray*}
\Delta_{in} (\Gamma)=\sum_{j=1}^i\Gamma_{ji}-\sum_{j=1}^{i-1}\Gamma_{j, i-1}
-\sum_{j=1}^nm_{jn}+\sum_{j=1}^{n-1}m_{j, n-1}
\end{eqnarray*}
\begin{eqnarray*}
[\Delta (\Gamma)]_n\equiv[\Delta_{in} (\Gamma)\cdots0]
\end{eqnarray*}
The  restriction chain (\ref{chain}) has an important consequence for the
unit tensor operators (\ref{utp}). It allows to define its action on a given
state in a recursive way~:
\begin{equation}
\left\langle\begin{array}{ccc}
 (\Gamma)_{n-1}\\\
[M]_n\\
 (M)_{n-1}\end{array}\right\rangle
\left|\begin{array}{ccc}
[m]_n\\\
[m]_{n-1}\\\
 (m)_{n-2}\end{array}\right\rangle =\sum_{ (\gamma)_{n-2}}
\left[\begin{array}{ccc}
 (\Gamma)_{n-1} & [m]_n\\\
[M]_n\\
 (\gamma)_{n-1} & [m]_{n-1}\end{array}\right] \left\langle\begin{array}{ccc}
 (\gamma)_{n-2}\\\
[M]_{n-1}\\
 (M)_{n-2}\end{array}\right\rangle \left|\begin{array}{ccc}
[m]_n +[\Delta (\Gamma)]_n\\\
[m]_{n-1}\\\
 (m)_{n-2}\end{array}\right\rangle
\label{2.9}
\end{equation}
where
\begin{eqnarray*}
 (\gamma)_{n-1}\equiv\left (
\begin{array}{cc}
[M]_{n-1}\\
 (\gamma)_{n-2}\end{array}\right)
\end{eqnarray*}
Here the $\left\langle\begin{array}{ccc}
 (\gamma)_{n-2}\\\
[M]_{n-1}\\
 (M)_{n-2}\end{array}\right\rangle$
 operator in the r. h. s. is the $U (n-1)$ unit tensor operator, acting on
 $| (m)_{n-1}\rangle$ which is the state from the irrep of $U (n-1)$ with
highest weight $[m]_{n-1}$ . The remaining part in (\ref{2.9})~:
\begin{eqnarray}
\left[\begin{array}{ccc}
 (\Gamma)_{n-1} & [m]_n\\\
[M]_n\\
 (\gamma)_{n-1} & [m]_{n-1}\end{array}\right]
\label{rto}
\end{eqnarray}
are c-numbers which depend on the $ (\Gamma)_{n-1},[M]_{n}, (\gamma)_{n-1}$
and on the first two rows of $| (m)\rangle$~: $[m]_n$ and $[m]_{n-1}$ . 
Such a decomposition has many interesting properties \cite{B} and it will
be useful in the following .

The non-zero matrix elements of the unit tensor operators can be
understood as the Clebsch-Gordan coefficients of $SU (n)$ . We can
couple two state vectors $| (m)\rangle$ and $| (M)\rangle$ from the
space of direct product of irreps $[m]_n$ and $[M]_n$ to obtain coupled
state vectors which are again the Gelfand basis vectors for an irrep of
$SU (n)$~:
\begin{equation}
\left|\left (\begin{array}{cc}
[m]+[\Delta (\Gamma )]\\
 (m^\prime)\end{array}\right); (\Gamma )\right\rangle=
\label{interpr}
\end{equation}
\begin{eqnarray*}
\sum_{ (M) (m)}
\left\langle \left (\begin{array}{cc}
[m]+[\Delta (\Gamma )]\\
 (m^\prime )\end{array}\right)\right|
\left\langle \begin{array}{ccc}
 (\Gamma )\\\
[M]\\
 (M)\end{array}\right\rangle \left| \begin{array}{cc}
[m]\\
 (m)\end{array}\right\rangle
\left| \begin{array}{cc}
[M]\\
 (M)\end{array}\right\rangle \times
\left| \begin{array}{cc}
[m]\\
 (m)\end{array}\right\rangle
\end{eqnarray*}

Because the operators $u_{ij}$ are   tensor operators in the
 fundamental representation, we  will need the explicit form of (\ref{utp})
only for  $[M]_n =[1 0 \cdots 0]$. 
For these tensor operators and for the corresponding numbers (\ref{rto}) in the
decomposition (\ref{2.9}) it is possible to introduce the simplified notations~:
\begin{mathletters}
 
\begin{eqnarray}
\left\langle\begin{array}{ccc}
 (\Gamma)_{n-1}\\\
[10\cdots 0]\\\
 (M)_{n-1}\end{array}\right\rangle \rightleftharpoons \left\langle
\begin{array}{cc}
i\\
j\end{array}\right\rangle
\label{2.12}
\end{eqnarray}
 
\begin{eqnarray}
 \left[\begin{array}{ccc}
 (\Gamma)_{n-1} & [m]_n\\\
[M]_n\\
 (M)_{n-1} & [m]_{n-1}\end{array}\right] \rightleftharpoons
\left[\begin{array}{cc}
i & [m]_n\\
j & [m]_{n-1}\end{array}\right]
\label{2.12a}
\end{eqnarray}
 
\end{mathletters}
 
The r. h. s. of (\ref{2.12}) denotes the (unique) tensor operator with
 upper and lower pattern defined by the shift (\ref{shift}) $\Delta_n (i),
\hskip 0.2 cm (\Delta_n (j))$ respectively . For $i \neq n$ we have~:
 
\begin{eqnarray}
\Delta_n (i)=[0\cdots 01 0\cdots 0]\hskip 0.3 cm,
\hskip 0.3 cm
\Delta_{n} (j)=[0\cdots 01 0\cdots 0]
\end{eqnarray}
 
where $1$ appears in the position $i (j)$. 
For $i=n$ we have~:
\begin{eqnarray}
\Delta_n (n)=[-1 -1 \cdots -1 0]
\hskip 0.3 cm ,\hskip 0.3 cm
\Delta_{n} (n)=[-1 -1\cdots-1 0 -1\cdots -1]
\end{eqnarray}
where $0$ appears in the position $j$. 
The numbers $\left[\begin{array}{cc}
i & [m]_n\\
j & [m]_{n-1}\end{array}\right]$ can be found explicitly \cite{B}~:
 
\begin{eqnarray}
\left[\begin{array}{cc}
i & [m]_n\\
j & [m]_{n-1}\end{array}\right]^2 =
\frac{\prod_{j^\prime \neq j}^{n-1} (p_{in}-p_{j^\prime, n-1})\prod_{i^\prime
 \neq i}^n (p_{j, n-1}-p_{i^\prime, n}+1)
}{\prod_{i^\prime \neq i}^n (p_{in}-p_{i^\prime, n})\prod_{j^\prime \neq j}
^{n-1} (p_{j, n-1}-p_{j^\prime, n-1}+1)
\label{2.13}}
\end{eqnarray}
where $p_{in}=m_{in}+n-i ; p_{nn}\equiv 0$. 
The coefficients in (\ref{2.13})
 are chosen in such a way that $u_{ij}$ operators correspond to the matrix 
elements of the unitary matrix~:
\begin{mathletters}
\begin{eqnarray}
\label{norm1}
\sum_j u_{jk}^{+}u_{ij}=\delta_{ik}
\end{eqnarray}
\begin{eqnarray}\sum_j u_{jk}u_{ij}^{+}=\hat{I}_j \delta_{ik}
\end{eqnarray}
\end{mathletters}
where $\hat{I}_j |(m)\rangle =0$ if $u_{ij}^{+}|(m)\rangle =0$ , otherwise 
$\hat{I}_j |(m)\rangle =|(m)\rangle$.

 In the Appendix 1 it is shown that the commutation
relations of the unit tensor operators in the fundamental representation
  coincide with the $B$ matrix relations (\ref{2.6}), (\ref{2.7}) for
the $u_{ij}$ operators. Therefore the explicit realization of the $u_{ij}$
operators is
\begin{eqnarray}
u_{ji}=\left\langle
\begin{array}{cc}
i\\
j\end{array}\right\rangle
\end{eqnarray}

\subsection {Combining the left and right theories. }

 After quantizing the chiral part separately we return now to our original model
 (\ref{act1}). As discussed in section \ref{2B}
the physical Hilbert space ${\cal H}_{phys}$ will be the direct product of
chiral Hilbert spaces on which we have imposed the "Gauss law"~:
 
\begin{equation}
\forall |\psi\rangle \in {\cal H}_{phys}\hskip 1 cm
 (\vec{p_l}-\vec{p_r})|\psi\rangle=0
\end{equation}
It means that~:
\begin{equation}
{\cal H}_{phys}=\bigoplus_{\{ \lambda\}}{\cal H}^l_{\lambda}
 \otimes {\cal H}^r_{\lambda}
\end{equation}
where $\{ \lambda\}$ is the set of all irreps of $SU (n)$ with highest weights
 $\lambda$ and  the left and  right irreps have the same highest weights. 
 Therefore
the states in the physical Hilbert space are labeled by $\lambda$  and the
magnetic quantum numbers $ (\gamma)_{n-1} , (\chi)_{n-1}$ . In the Gelfand pattern
notation the states in ${\cal H}_{phys}$ can be denoted by a  double Gelfand
pattern~:
\begin{equation}
\left|\begin{array}{ccc}
 (\gamma)_{n-1}\\\
[m]_n\\
 (\chi)_{n-1}\end{array}\right\rangle
\end{equation}
where the upper (lower) parts of this double pattern designate the left
 (right) state respectively. It is isomorphic to the Hilbert space  in the
standard treatment discussed in the section \ref{2A}, if the matrix
elements of the representation are labeled by the highest weight $[m]_n$
and the magnetic quantum numbers $ (\gamma)_{n-1} , (\chi)_{n-1}$ .

We discuss now the physical operators in the theory. All physical operators
must commute
with $\vec{\phi}$. The operator corresponding to the $g$-operator in the
 initial notations is given by~:
\begin{equation}
g_{ij}=\sum_{k=1}^n u_{ik}v_{kj}
\label{gdef}
\end{equation}
It is easy to check that~:
\begin{eqnarray}
[g_{ij}, (\vec{p_l}-\vec{p_r})]=0
\end{eqnarray}
 
Therefore we can consistently restrict  $g$ to the physical Hilbert space. 
We check now that in the restricted Hilbert space~:
\begin{equation}
[g_1, g_2]=0
\end{equation}
Indeed it follows from (\ref{2.6}), (\ref{2.7}) that~:
\begin{eqnarray}
g_{ij}g_{kl}=\sum_{qs}u_{iq}v_{qj}u_{ks}v_{sl}=
\end{eqnarray}
\begin{eqnarray*}
\sum_{qs} (u_{kq}u_{is}B^{l1}_{qs}+u_{ks}u_{iq}B^{l2}_{qs})
 (B^{r1}_{qs}v_{ql}v_{sj}+B^{r2}_{qs}v_{sl}v_{qj})=
\end{eqnarray*}
\begin{eqnarray*}
\sum_{qs} (u_{kq}u_{is}v_{ql}v_{sj} (B^{l1}_{qs}B^{r1}_{qs}+
B^{l2}_{qs}B^{r2}_{qs})+u_{kq}u_{is}v_{sl}v_{qj} (B^{l1}_{qs}B^{r2}_{qs}+
B^{l2}_{sq}B^{r1}_{sq}))
\end{eqnarray*}
where e. g. 
\begin{eqnarray*}
B^{l1}_{qs}=\frac{\hbar}{p_{lq} -p_{ls}}, \hskip 1 cm
B^{l2}_{qs}=\sqrt{1-\left ( \frac{\hbar}{p_{lq} -p_{ls}}\right)^2}
\end{eqnarray*}
and  similar expressions for $B^{r1, 2}_{qs}$. The definition of the physical
Hilbert space and the antisymmetry of $B^{ (l, r)1}_{qs}$ with
respect to the permutation $ (qs)\rightarrow (sq)$ prove the statement. 
We can calculate explicitly the action of the $g$ operator on the ground
state using (\ref{gdef})~:
\begin{equation}
g_{ij}\left|\begin{array}{ccc}
 (0)_{n-1}\\\
[0]_n\\
 (0)_{n-1}\end{array}\right\rangle=\left|\begin{array}{ccc}
 (\gamma)_{n-1}\\\
[10 \cdots 0]_n\\
 (\chi)_{n-1}\end{array}\right\rangle
\label{gact}
\end{equation}
where
\begin{eqnarray*}
\Delta_{n} (\gamma)=[0 \cdots 01^i 0 \cdots 0]
\end{eqnarray*}
 
\begin{eqnarray*}
\Delta_{n} (\chi)=[0 \cdots 01_j 0 \cdots 0]
\end{eqnarray*}
Equation (\ref{gact}) reproduces the result in the standard treatment \cite{M}. 
The Hamiltonian of the full system is given by~:
\begin{equation}
{\cal H}=\frac {i}{2}\sum_{k=1}^rp_k p_k-\hbar^2 tr (H^2H^2) \hskip 0.7 cm ,
\hskip 0.7 cm {\mbox where} \hskip 0.5 cm  p_k=\frac{1}{2} (p_k^l+p_k^r)
\label{gham}
\end{equation}
and  it  coincides with the second-order Casimir operator. Using
 (\ref{gact}) and (\ref{gham}) we recover the propagator (\ref{prop}).

\section{Chiral Quantization of the $WZW$ model}
 
\subsection{Classical analysis in Chiral Variables}
 
We return now to our main problem , the chiral quantization of the $WZW$
model. Let us consider the group-valued field $g (\tau ,x)$ , where $x$ is the
spatial coordinate for a circle with unit radius , and $g (\tau ,x+2\pi)=
g (\tau ,x)$. Let $\psi$ denote the longest root. The action for the $WZW$
model can be written as \cite{W}~:
 
\begin{equation}
S[g]=-\frac{k\psi^2}{16\pi}\int_{{\cal{M}}}tr (\partial_{\mu}g
\partial^{\mu}g^{-1})d^2 x+\Gamma [g]
\label{1.30}
\end{equation}
where  $\Gamma [g]$ is the topological term~:
\begin{equation}
\Gamma [g]=\frac{k\psi^2}{24\pi}\int_{{\cal{B}}}d^3 X\epsilon^{\alpha \beta \gamma}
tr (\tilde{g}^{-1}\partial_{\alpha}\tilde{g}\tilde{g}^{-1}\partial_{\beta}
\tilde{g}\tilde{g}^{-1}\partial_{\gamma}\tilde{g})
\label{1.31}
\end{equation}
Here $\cal{B}$ is the three dimensional domain whose boundary is the cylinder
$\cal{M}$ and $\tilde{g} (\tau ,x,y)\in G$ is the map from $\cal{B}$ to $G$
such that $\tilde{g} (\tau ,x,1)=g (\tau ,x)$. The singlevaluedness of the
probability amplitude requires $k$ to be an integer. 
 
 The equations of motion following from the action (\ref{1.30}) are~:
\begin{eqnarray*}
\partial_{+} (g^{-1}\partial_{-}g) = \partial_{-} (\partial_{+}gg^{-1}) = 0
\hskip 0.4 cm , \hskip 0.4 cm  x^{\pm} \equiv \tau \pm x
\end{eqnarray*}
 The general solution of these equations is~:
\begin{eqnarray*}
g (\tau , x)=\tilde{u} (x^{+}) (\exp{2i\vec{p}\vec{H} (x^{+}+x^{-})})\tilde{v} (x^{-})
\end{eqnarray*}
 This solution is invariant under the transformation~:
 \begin{equation}
\tilde{u}\rightarrow \tilde{u}h  ,\hskip 1 cm  \tilde{v}\rightarrow h^{-1} tilde{v} \hskip 0.4 cm  ,
 \hskip 0.4 cm  h \in T \label{3.1}
\end{equation}
 like the solution of the finite-dimensional model. 
 
 Using the fact that  any solution is completely defined by the
 given $g (\tau , x)|_{\tau=0}$ and
 $\partial_{\tau}g (\tau , x)|_{\tau =0}$
let us change variables in analogy with what we did in section 2 \cite{sm}~:
\begin{equation}
g (\tau , x)|_{\tau=0}=\tilde{u}\tilde{v}
\end{equation}
\begin{eqnarray*}
g^{-1}\partial_{\tau}g|_{\tau=0}=\tilde{v}^{-1}\tilde{u}^{-1}\partial_{x}\tilde{u}\tilde{v} +
 4i\tilde{v}^{-1}\vec
{p}\vec{H}\tilde{v} - \tilde{v}^{-1}\partial_{x}\tilde{v}
\end{eqnarray*}
where the periodic variables $\tilde{u}, \tilde{v}$ belong to the loop group $ LG$ and we used
the following identity~:
\begin{equation}
\partial_{\tau}\tilde{u}|_{\tau =0}=\partial_{x}\tilde{u} \hskip 0.5 cm ; \hskip 0.5 cm
 \partial_{\tau}\tilde{v}|_{\tau =0}=-\partial_{x}\tilde{v}
\end{equation}
 In these variables the action takes the form~:
\begin{equation}
S=\frac{k\psi^2}{8\pi} \{ \int_{\cal M} tr (\tilde{u}^{-1}\partial_x \tilde{u}\tilde{u}^{-1}
\partial_{t}\tilde{u} -\tilde{v}^{-1}\partial_x \tilde{v}\tilde{v}^{-1}\partial_{t}\tilde{v} +
\end{equation}
\begin{eqnarray*}
+4i\vec{p}\vec{H} (\tilde{u}^{-1}\partial_{t}\tilde{u} + \partial_{t}\tilde{v}\tilde{v}^{-1}-\tilde{u}^{-1}
\partial_x \tilde{u} + \partial_x \tilde{v}\tilde{v}^{-1})
\end{eqnarray*}
\begin{eqnarray*}
 +8 (\vec{p}\vec{H})^2- (\tilde{u}^{-1}\partial_x \tilde{u})^2- (\tilde{v}^{-1}\partial_x \tilde{v})^2)dxdt
\} +\Gamma[\tilde{u}]+\Gamma[\tilde{v}]
\end{eqnarray*}
where $\Gamma[\tilde{u}]$ and $\Gamma[\tilde{v}]$ are the topological terms. 
 
This action and the relevant symplectic  form are degenerate due to the
 invariance (\ref{3.1}). The situation is completely analogous to the "toy"
model (\ref{ambig}) . In a similar way we introduce separate $\vec{p}_l$
and $\vec{p}_r$ and the constraints (\ref{constr}). The new action is~:
 \begin{equation}
S=\frac{k\psi^2}{8\pi} \{ \int_{\cal M} tr (\tilde{u}^{-1}\partial_x \tilde{u}\tilde{u}^{-1}
\partial_{t}\tilde{u} -\tilde{v}^{-1}\partial_x \tilde{v}\tilde{v}^{-1}\partial_{t}\tilde{v} +4i\vec{p_l}\vec{H}
 (\tilde{u}^{-1}\partial_{t}\tilde{u}-\tilde{u}^{-1} \partial_x \tilde{u})+
\end{equation}
\begin{eqnarray*}
4i\vec{p_r}\vec{H} (\partial_{t}\tilde{v}\tilde{v}^{-1} + \partial_x \tilde{v}\tilde{v}^{-1})+2 ( (\vec{p_l}
+\vec{p_r})\vec{H})^2- (\tilde{u}^{-1}\partial_x \tilde{u})^2- (\tilde{v}^{-1}\partial_x \tilde{v})^2)dxdt\}
+\end{eqnarray*}
\begin{eqnarray*}
+\Gamma[\tilde{u}]+\Gamma[\tilde{v}]
\end{eqnarray*}
The corresponding symplectic  form is~:
\begin{equation}
\label{symplectic}
\Omega =\frac{k\psi^2}{8\pi}\int tr (\tilde{u}^{-1}\delta
\tilde{u}\partial_x (\tilde{u}^{-1}\delta \tilde{u})
 + 4i\vec{p_l}\vec{H}\tilde{u}^{-1}\delta \tilde{u}\tilde{u}^{-1}\delta \tilde{u} - d\vec{p_l}\vec{H}\tilde{u}^{-1}
\delta \tilde{u} -
\end{equation}
\begin{eqnarray*}
 -\delta \tilde{v}\tilde{v}^{-1}\partial_x (\delta \tilde{v}\tilde{v}^{-1}) + 4i\vec{p_r}\vec{H}\delta
 \tilde{v}\tilde{v}^{-1}\delta \tilde{v}\tilde{v}^{-1} - d\vec{p_r}\vec{H}\delta \tilde{v}\tilde{v}^{-1})
\end{eqnarray*}
It is convenient to introduce the new variables~:
\begin{equation}
u(x)=\tilde{u}(x)\exp{2i\vec{p}_l \vec{H}x} \hskip 1.5cm , \hskip 1.5cm
v(x)=\tilde{v}(x)\exp{-2i\vec{p}_r \vec{H}x}
\end{equation}
The symplectic form (\ref{symplectic}) is non-degenerate and we can obtain the Poisson brackets~:
\begin{equation}
\{ u (x), \vec{p_l} \} = \frac{\beta}{2}u (x)\vec{H} \hskip 0.4 cm  ,
 \hskip 0.4 cm  \{ v (x), \vec{p_r} \} = \frac{\beta}{2} \vec{H} v (x)
\label{3.2}
\end{equation}
\begin{eqnarray*}
\{ u_1 (x), u_2 (y) \} = u_1 (x)u_2 (y)R_l (x-y)
\end{eqnarray*}
\begin{eqnarray*}
\{ v_1 (x), v_2 (y) \} =R_r (x-y) v_1 (x)v_2 (y)
 \end{eqnarray*}
where
\begin{mathletters}
\begin{equation}
R_l (x) =\frac{\beta}{2}\eta (x)\sum_{j=1}^{r}H^j \otimes H^j +
i\frac{\beta}{2}\sum_{\alpha}\frac{1}{\sin (\alpha \cdot p_l)}e^{-i\alpha
 \cdot p_l \eta (x)}E_{\alpha}  \otimes E_{-\alpha}
\end{equation}
\begin{eqnarray}
R_r (x) =\frac{\beta}{2}\eta (x)\sum_{j=1}^{r}H^j \otimes H^j
-i\frac{\beta}{2}\sum_{\alpha}\frac{1}{\sin (\alpha \cdot p_r)}e^{-i\alpha
 \cdot p_r \eta (x)}E_{\alpha}  \otimes E_{-\alpha}
\end{eqnarray}
\end{mathletters}
\begin{eqnarray*}
\beta =\frac{4\pi}{\psi^2 k}
\end{eqnarray*}
and $\eta (x)=2[\frac{x}{2\pi}]+1 \hskip 0.2 cm  , \hskip 0.2 cm  [x]$ denotes
 the  maximal integer , less then $x$. The Poisson brackets between left and
 right variables vanish. Since the Poisson brackets (\ref{3.2}) are derived from
a non-degenerate lagrangian they fulfil automatically the Jacobi identities. 
 
 From these Poisson brackets it is possible to derive the classical
 transformation properties of $u (x)$ and $v (x)$ under Kac-Moody symmetry
 action~:
\begin{mathletters}
\begin{equation}
\{ J_l^a (x), u (y)\} = it^a u (y)\delta (x-y)
\end{equation}
\begin{eqnarray}
\{ J_r^a (x), v (y)\} = iv (y)t^a \delta (x-y)
\end{eqnarray}
\end{mathletters}
where the Kac-Moody currents are given by~:
\begin{equation}
J_l (x)= \frac{ik}{4\pi}\partial_x uu^{-1} \hskip 0.3 cm  , \hskip 0.3 cm
J_l (x)= \frac{ik}{4\pi}v^{-1}\partial_x v
\end{equation}

\subsection{The quantization of the chiral theory}

 We quantize these Poisson brackets , according
\cite{Blo}\cite{Fad}\cite{Sch}\cite{G1}. In the following
we will map  the cylinder to the complex plane $e^{ix}\rightarrow z$ . 
 From now on we will consider only the $\hat{SU} (n)_k$ $WZW$ model , with the
$u$ and $v$ operators in the fundamental representation. The algebra of the
quantum operators is~:
\begin{mathletters}
\label{3.3}
\begin{equation}
[\vec{p}_l , u (z)]=-\frac{\beta}{2}\hbar u (z)\vec{H} \hskip 0.3 cm  , \hskip
 0.3 cm [\vec{p}_l , v (\bar{z})]=-\frac{\beta}{2}\hbar \vec{H}v (\bar{z})
\label{3.3a}
\end{equation}
\begin{equation}
u_1 (z_1 ) u_2 (z_2 ) = u_2 (z_2) u_1 (z_1 ) B_l (\frac{z_1}{z_2})
\hskip 0.3 cm  , \hskip 0.3 cm
v_1 (\bar{z}_1 ) v_2 (\bar{z}_2 ) = B_r (\frac{\bar{z}_1}{\bar{z}_2})v_2
 (\bar{z}_2 ) v_1 (\bar{z}_1 )\label{3.3b}
\end{equation}
\begin{equation}
J^a_l (z_1 )u (z_2 )=\frac{it^a u (z_2 )}{z_1 -z_2 } +\cdots
\hskip 0.3 cm  , \hskip 0.3 cm
J^a_r (\bar{z}_1 )v (\bar{z}_2 ) =\frac{i v (\bar{z}_2 )t^a }{\bar{z}_1 -\bar{z}_2}
+ \cdots
\end{equation}
\end{mathletters}
where $B_{l, r} (z)$ are the braiding matrices , which can be obtained by
exponentiating the classical $R-$matrix , like (\ref{2.7}). 
 It is enough
to consider only the left-invariant theory , because the right one is
completely analogous. 
 
The explicit expression for $B_l (z)$ is~:
\begin{equation}
\label{3.4}
B_l (z)=q^{ (\frac{n-1}{n})\eta (arg (z))}\left ( 1 \otimes 1 -
\sum_{\alpha \in \Phi}
E_{\alpha}E_{-\alpha} \otimes E_{-\alpha}E_{\alpha} \right)
\end{equation}
\begin{eqnarray*}
+q^{-\frac{1}{n}\eta (arg (z))}\sum_{\alpha \in \Phi}\cos\theta (\alpha \cdot p_l)
E_{\alpha}E_{-\alpha}\otimes E_{-\alpha}E_{\alpha}\hskip 2.5 cm  q\equiv
 e^{\frac{i\pi}{k+n}}
\end{eqnarray*}
\begin{eqnarray*}
-q^{-\frac{1}{n}\eta (arg (z))}\sum_{\alpha \in \Phi}e^{-i\alpha \cdot p_{l}}
\sin\theta (\alpha \cdot p_l)E_{\alpha}\otimes E_{-\alpha}
\end{eqnarray*}
where
\begin{eqnarray*}
\sin\theta (\alpha \cdot p_l)\equiv \frac{\sin\frac{\pi}{k+n}}{\sin \alpha \cdot p_l}
\end{eqnarray*}
The expression for $\beta$ including the quantum correction is~:
\begin{eqnarray}
\beta \hbar =\frac{2i\pi}{k+n}
\end{eqnarray}
The braiding matrix can be related to the Racah matrix of $U_{2q} (SL (n))$
 in the fundamental representation. 
 
 We  assume  that the Hilbert space of the system is~:
\begin{equation}
\cal H =\bigoplus_{\{ \lambda \}}\cal H_{\lambda}
\end{equation}
where $\{ \lambda \}$ denotes the set of
all integrable representations of $\hat{SU} (n)_{k}$ . 
The integrability condition for the $\hat{SU} (n)_k$ \cite{GW} is~:
\begin{equation}
k\geq m_{1n}
\label{int}
\end{equation}
We will find an explicit representation of all chiral operators on this
Hilbert space. 
Because the $p^i_l$ operators
 commute with all Kac-Moody currents , they depend only on the highest
 weights of the representations and we can define~:
\begin{equation}
p^i_l |[m]_n, (m)\rangle =\frac{\hbar}{k+n} (m_{in}+n-i)|[m]_n , (m)\rangle
 \label{3.5}
\end{equation}
where $[m]_n$ denotes the highest weight of some integrable representation
 and $ (m)$ denotes all other quantum numbers characterizing the state. The
commutation relations of $u_{ij} (z)$ and $J^{ (l)a}_n$ imply  that
$u_{ij} (z)$ is a primary field in the fundamental representation. For its
matrix elements between the zero level states one has~:
\begin{equation}
\Psi_1 =\langle (\tilde{m})^{\infty}_n | u_{ij} (z) | (m)^0_n \rangle
 =\frac{Inv (V_{\tilde{m}}^{\ast}\otimes V\otimes V_m)}{z^{-h (\tilde{m})
+h+h (m)}} \label{3.6}
\end{equation}
where $Inv (V_{\tilde{m}}^{\ast}\otimes V\otimes V_m)$ is a $SU (n)$
 invariant functional on $V_{\tilde{m}}^{\ast}\otimes V\otimes V_m$ ;
$V_{\tilde{m}}, V_m$ are the irreps of $SU (n)$ with the highest weights
$[\tilde{m}]_n$ , $[m]_n$ respectively and $V$ is the fundamental
representation. The $h (m)$ is the conformal dimension of the operator~:
\begin{equation}
h (m) = \frac{C_2 (m)}{k+n}
\end{equation}
and $C_2 (m)$ is  the value of the second-order Casimir operator. 
All other matrix elements can be computed using the chiral symmetry. 
Equations (\ref{3.3a}), (\ref{3.5}) imply  that $\Psi_1 (z)$ is not equal to
zero only if the highest weights are related by~:
\begin{equation}
[\tilde{m}]_n = [m]_n + \Delta_n (j)
\label{diff1}
\end{equation}
 Equation (\ref{diff1}) and the interpretation of the matrix elements of
the unit tensor operators (\ref{interpr}) means that we can write $\Psi_1 (z)$
as follows~:
\begin{equation}
\Psi_1 (z) = \frac{\langle (\tilde{m})_n | \gamma_{ij}| (m)_n \rangle}
{z^{-h (\tilde{m})+h+h (m)}}C_{[m]_n , j}
\label{diff2}
\end{equation}
where $\gamma_{ij}$ is the unit tensor operator , defined earlier (\ref{2.12}),
 (\ref{2.13}) ,
and $C_{[m]_n , j}$ is a constant depending only on $j$ and the
highest weight $[m]_n$. The conditions (\ref{diff1}), (\ref{diff2}) mean that
the right index of the chiral vertex operator $u_{ij} (z)$ determines the
difference between integrable highest weights $[m]_n$ and $[\tilde{m}]_n$
like the right index of the unit tensor operators $\gamma_{ij}$. Using
the projected vertex operators $\Phi_{\tilde{m}m}^k (z)$ \cite{TK}
\cite{MS}
one can write the $u_{ij} (z)$ as~:
\begin{eqnarray}
u_{ij} (z)= \sum_{[\tilde{m}], [m]}\Phi_{\tilde{m}m}^i (z)C_{[m]_n , j}\delta_
{[\tilde{m}]_n , [m_n +\Delta_n (j)]}
\label{pvo}
\end{eqnarray}
where the $\Phi_{\tilde{m}m}^i$ is in the fundamental representation with
magnetic quantum number $i$ and the sum is over all integrable highest weights
$[\tilde{m}]_n , [m]_n$. 
 
The  commutation relations (\ref{3.3b}) imply  restrictions on the constants
$C_{[m]_n , j}$ . To determine these restrictions we need to find the matrix
elements of the product of the two $u$ operators between  arbitrary states. 
However it is enough to consider only the matrix elements of this product
between the zero level states, since the $B$ matrix (\ref{3.4}) commutes with
the Kac-Moody currents~:
\begin{equation}
\Psi_2 (z_1 ,z_2 ) = \langle (\tilde{m})^{\infty}_n |u_{ij} (z_1 )u_{kl} (z_2 )
| (m)^0_n \rangle
\label{uproduct}
\end{equation}
 In the discussion of the
tensorial properties of (\ref{uproduct}) we can use the results for the
"toy model" because the level zero subspace in the representation of
$\hat{SU} (n)_k$ is equivalent to the irreducible representation of  $SU (n)$. 
 To determine the $z_1 ,z_2$ dependence of (\ref{uproduct}) it is convenient
to use the KZ equations \cite{KZ} in the form proposed in \cite{R}. 
 
The tensorial structure of (\ref{uproduct}) is given by the condition~:
 
\begin{equation}
\Psi_2 (z_1 ,z_2 ) \in Inv (V^{\ast}_{\tilde{m}}\otimes V\otimes V\otimes V_{m})
\end{equation}
 The dimension of $Inv (V^{\ast}_{\tilde{m}}\otimes V\otimes V\otimes
V_{m})$ is different from zero only if~:
\begin{equation}
[\tilde{m}]_n = [m]_n + \Delta_n (j) + \Delta_n (l)
\end{equation}
We will treat separately the cases~: $j \neq l$ and $j=l$. 
 
Let us first consider the $\hskip 0.3 cm  j \neq l$ case. The space of
invariant couplings for the given
$| (\tilde{m})\rangle \in V_{\tilde{m}}$ , $| (m)\rangle \in V_{m}$
is two dimensional. It is convenient to choose as a basis in this space the
following vectors~:
\begin{equation}
T_1 =\langle (\tilde{m})|\gamma_{ij} \gamma_{kl} | (m)\rangle \hskip 0.3 cm  ;
\hskip 0.3 cm  T_2 =\langle (\tilde{m})|\gamma_{kj} \gamma_{il} | (m)\rangle
\end{equation}
where $\gamma_{pq}$ is the unit tensor operator. 
 
In these notations $\Psi_2 (z_1 , z_2 )$ takes the form~:
\begin{equation}\Psi_2 (z_1 , z_2 )=T_1 \psi_1 (z_1 , z_2 )+T_2 \psi_2 (z_1 ,
z_2 )\end{equation}
where $\psi_{1, 2}$ depends only on $j, l$ and the highest weights $
[\tilde{m}]_n , [m]_n$. Following \cite{R} we can write the KZ equations for
$\Psi_2 (z_1 , z_2 )$ in the form~:
\begin{mathletters}
\begin{equation}
 (k+n)\frac{\partial \Psi_2}{\partial z_1}=\left (\frac{t_1 \otimes t_2}
{z_1 -z_2}+\frac{t_1 \otimes t_{m}}{z_1 }\right )\Psi_2
\end{equation}
\begin{equation}
 (k+n)\frac{\partial \Psi_2}{\partial z_2}=\left (\frac{t_2 \otimes t_1}
{z_2 -z_1}+\frac{t_2 \otimes t_{m}}{z_2 }\right )\Psi_2
\end{equation}
\end{mathletters}
where $t_i$ are the generators of $SU (n)$ in the $i$ representation. 
 
Using the properties of the unit tensor operators $\gamma_{pq}$ we get (see
  Appendix 2)~:
\begin{mathletters}
\label{3.7}
\begin{equation}
 (t_1 \otimes t_2 )T_1 =T_2 -\frac{1}{n}T_1 \hskip 0.3 cm  , \hskip 0.3 cm
 (t_1 \otimes t_2 )T_2 =T_1 -\frac{1}{n}T_2 \label{3.7a}
\end{equation}
\begin{equation}
 (t_1 \otimes t_{m} )T_1 =-T_2 +\frac{1}{n}T_1 +\frac{1}{2} (C_2 ([\tilde{m}
]_n )-C_2 ([m]_n +\Delta_n (l))-C_2 )T_1 \label{3.7b}
\end{equation}
\begin{equation}
 (t_2 \otimes t_{m} )T_2 =-T_1 +\frac{1}{n}T_2 +\frac{1}{2} (C_2 ([\tilde{m}
]_n )-C_2 ([m]_n +\Delta_n (l))-C_2 )T_2 \label{3.7c}
\end{equation}
\begin{equation}
 (t_2 \otimes t_{m} )T_1 =\frac{1}{2} (C_2 ([m]_n +\Delta_n (l))-C_2 ([m]_n )
-C_2 )T_1 \label{3.7d}
\end{equation}
\begin{equation}
 (t_1 \otimes t_{m} )T_2 =\frac{1}{2} (C_2 ([m]_n +\Delta_n (l))-C_2 ([m]_n )
-C_2 )T_2 \label{3.7e}
\end{equation}
\end{mathletters}
After the substitution~:
\begin{equation}
\psi_{1, 2} (z_1 , z_2 )= (z_1 z_2 )^{\frac{C_2
 ([\tilde{m}]_n) +C_2 ([m]_n )-2C_2}{4 (k+n)}}\chi_{1, 2} (z)
\end{equation}
 where
 $z=\frac{z_2}
{z_1}$ we get the system of ordinary differential equations~:
\begin{equation}
\label{eq}
\frac{d}{dz}\left (\begin{array}{rr}
\chi_1 (z)\\
\chi_2 (z)\end{array}\right) =
\end{equation}
\begin{eqnarray*}
\left\{\frac{1}{z}\left (\begin{array}{cc}
-\frac{1}{2} (\mu +\beta) & -\gamma \\
0 & \frac{1}{2} (\mu +3\beta)\end{array}\right) +
\frac{1}{1-z}\left ( \begin{array}{cc}
\beta & -\gamma \\
-\gamma & \beta \end{array}\right) \right\}
 \left ( \begin{array}{rr}
\chi_1 (z) \\
\chi_2 (z) \end{array}\right)
\end{eqnarray*}
where
\begin{eqnarray*}
\mu =\frac{C_2 ([\tilde{m}]_n)+C_2 ([m]_n)-2C_2 ([m]_n +\Delta_n (l))}{2 (k+n)}
-\frac{1}{n (k+n)}
\end{eqnarray*}
\begin{eqnarray*}
\beta =\frac{1}{n (k+n)}\hskip 0.3 cm ; \hskip 0.3 cm  \gamma =\frac{1}{k+n}
\end{eqnarray*}
Using the explicit expression for $C_2 ([m]_n)$~:
\begin{eqnarray*}
C_2 ([m]_n )=\sum_{i=1}^{n-1} (m_{in})^2 -\frac{1}{n} (\sum_{i=1}^{n-1}m_{in})
^2 +\sum_{i<j} (m_{in}-m_{jn})
\end{eqnarray*}
we obtain~:
\begin{equation}
\mu =\frac{m_{jn}-m_{ln}+l-j}{k+n}
\end{equation}
The equations (\ref{eq}) are of Gaussian type. Their general solution is~:
\begin{mathletters}
\begin{equation}
\chi_1^+ =C^+ (1-z)^{-\beta +\gamma}z^{-\frac{1}{2} (\mu -\beta)}
F (\gamma , 1+\gamma -\mu , 1-\mu , z)
\end{equation}
\begin{equation}
\chi_1^- =C^- (1-z)^{-\beta +\gamma}z^{\frac{1}{2} (\mu +\beta)}
F (1+\gamma , \gamma +\mu , 1+\mu , z)
\end{equation}
\begin{equation}
\chi_2^+ =C^+ \frac{\gamma}{\mu -1} (1-z)^{-\beta +\gamma}z^{1-\frac{1}{2}
 (\mu -\beta)}F (1+\gamma , 1+\gamma -\mu , 2-\mu , z)
\end{equation}
\begin{equation}
\chi_2^- =-C^- \frac{\mu}{\gamma} (1-z)^{-\beta +\gamma}z^{\frac{1}{2}
 (\mu +\beta)}F (\gamma , \gamma +\mu , \mu , z)
\end{equation}
\end{mathletters}
where $F (\alpha , \beta , \gamma , z)$ are the hypergeometric functions. 
The functional form of these solutions coincides with the one obtained from the
fusion rules argumentation \cite{G}. 
Let us notice also  that if we  take
\begin{eqnarray*}
[m]_n =[11\cdots 1\cdots 10]
\end{eqnarray*}
\begin{eqnarray*}
[\tilde{m}]_n =[10\cdots 0\cdots 00] \hskip 0.4 cm  ; \hskip 0.4 cm
j=1\hskip 0.2 cm  , \hskip 0.2 cm  l=n
\end{eqnarray*}
we  obtain the KZ solution \cite{KZ}. 
 
From the expression for the matrix elements of $u_{ij} (z)$ between the zero
level states
we expect that for $z \rightarrow 0$~:
\begin{equation}
\lim_{z \rightarrow 0} \Psi_2 (z)=T_1 C_{[m]_n +\Delta_n
 (l), j}C_{[m]_n , l}z^{C_2 ([m]_n +\Delta_n (l))-C_2 ([m]_n)
-C_2}
\end{equation}
\begin{eqnarray*}
+O (z^{C_2 ([m]_n +\Delta_n (l))-C_2 ([m]_n) -C_2 +1})
\end{eqnarray*}
It means that~:
\begin{equation}
\Psi_2 (z_1 , z_2)= (z_1 z_2)^{\frac{C_2 ([\tilde{m}]_n )+C_2 ([m]_n )-2C_2}
{4 (k+n)}} (T_1 \chi_1^+ (z)+T_2 \chi_2^+ (z)) \label{3.8}
\end{equation}
 For $C^+$ we get~:
\begin{equation}C^+ =C_{[m]_n +\Delta_n (l), j}\times C_{[m]_n, l}\end{equation}
We can also write down the expressions for~:
\begin{mathletters}
\begin{equation}
\Phi_2 (z_1 , z_2)=\langle (\tilde{m})^{\infty}|u_{kj} (z_1)u_{il} (z_2)|
 (m)^0\rangle = (z_1 z_2)^{\Delta} (T_2 \chi_1^+ (z)+T_1 \chi_2^+ (z))
\end{equation}
\begin{equation}
\Xi_2 (z_1 , z_2)=\langle (\tilde{m})^{\infty}|u_{kl} (z_1)u_{ij} (z_2)| (m)^0
\rangle = (z_1 z_2)^{\Delta} (T_3 \xi_1^+ (z)+T_4 \xi_2^+ (z))
\end{equation}
\end{mathletters}
where $\Delta=\frac{C_2 ([\tilde{m}]_n)+C_2 ([m]_n)-2C_2}{4 (k+n)}$ .

The tensorial part of the solution $\Xi (z_1 , z_2 )$ is given by~:
\begin{mathletters}
\begin{equation}
T_3 =\langle (\tilde{m})|\gamma_{kl}\gamma_{ij}| (m)\rangle
\end{equation}
\begin{equation}
T_4 =\langle (\tilde{m})|\gamma_{kj}\gamma_{il}| (m)\rangle
\end{equation}
\end{mathletters}
 The $z-$dependence of $\Xi (z_1 , z_2)$  can be obtained simply by  changing
 the sign of $\mu $ in (\ref{3.8})~:
\begin{mathletters}
\begin{equation}
\xi_1 (z)=\tilde{C} (1-z)^{-\beta +\gamma}z^{\frac{1}{2} (\beta +\mu)}
F (\gamma , 1+\gamma +\mu , 1+\mu , z)
\end{equation}
\begin{equation}
\xi_2 (z)=-\tilde{C}\frac{\gamma}{1+\mu} (1-z)^{-\beta +\gamma}z^{1+
\frac{1}{2} (\beta +\mu)}F (1+\gamma , 1+\gamma +\mu , 2+\mu , z)
\end{equation}
\end{mathletters}
The behavior of the $\xi_{1, 2} (z)$ near the point $z=0$ implies that~:
\begin{equation}
\tilde{C}=C_{[m]_n +\Delta_n (j), l}\times C_{[m]_n , j}
\end{equation}
 
It can be shown (see  Appendix 3) that~:
\begin{mathletters}
\label{3.9}
\begin{equation}
\xi_1 (z)=-\frac{\tilde{C}}{C^- } (\frac{\gamma^2}{\mu^2 -\gamma^2}\chi_1^-
 (z)+\frac{\gamma \mu}{\mu^2 -\gamma^2}\chi_2^- (z)) \label{3.9a}
 \end{equation}
\begin{equation}
\xi_2 (z)=-\frac{\tilde{C}}{C^- } (\frac{\gamma \mu}{\mu^2 -\gamma^2}\chi_1^-
 (z)+\frac{\gamma^2 }{\mu^2 -\gamma^2}\chi_2^- (z))\label{3.9b}
 \end{equation}
\end{mathletters}
and we  conclude that the $\chi_{1, 2}^- (z)$ solution of the  KZ equations
 corresponds to the second channel of reaction. Using the analytic
properties of the hypergeometric functions\cite{Sp} we obtain~:
\begin{mathletters}
\begin{equation}
\chi_1^+ (\frac{1}{z})=\frac{C^+}{C^-}e^{-i\pi \beta}\frac{\gamma
\Gamma (1-\mu)\Gamma (-\mu)}{\Gamma (1-\mu -\gamma)\Gamma (1-\mu +\gamma)}
\chi^-_2 (z)+e^{-i\pi (\beta -\mu )}\frac{\sin\pi \gamma}{\sin\pi \mu}
\chi_2^+ (z)
\end{equation}
\begin{equation}
\chi_2^+ (\frac{1}{z})=\frac{C^+}{C^-}e^{-i\pi \beta}\frac{\gamma
\Gamma (1-\mu)\Gamma (-\mu)}{\Gamma (1-\mu -\gamma)\Gamma (1-\mu +\gamma)}
\chi^-_1 (z)+e^{-i\pi (\beta -\mu )}\frac{\sin\pi \gamma}{\sin\pi \mu}
\chi_1^+ (z)
\end{equation}
\end{mathletters}
The path of the analytic continuation is such that $-z=e^{i\pi \eta}z$,
 where $\eta \equiv \eta (arg\frac{z_1}{z_2})=1$. 
From (\ref{2.6}), (\ref{2.7}) one can
 derive the following connection between the invariant tensors $T_i$~:
\begin{mathletters}
\begin{equation}
T_3 =T_1 \frac{\mu}{\sqrt{\mu^2 -\gamma^2}}-T_2 \frac{\gamma}
{\sqrt{\mu^2 -\gamma^2}}
\end{equation}
\begin{equation}
T_4 =T_2 \frac{\mu}{\sqrt{\mu^2 -\gamma^2}}-T_1 \frac{\gamma}
{\sqrt{\mu^2 -\gamma^2}}
\end{equation}
\end{mathletters}
 
Combining these results with the commutation relations (\ref{3.7})~:
\begin{eqnarray*}
\Psi_2 (\frac{1}{z})=B_1 ([m]_n )\Phi_2 (z)+B_2 ([m]_n )\Xi_2 (z)
\end{eqnarray*}
\begin{eqnarray*}
{\rm where} \hskip 0.3 cm
B_1 ([m]_n)=e^{-i\pi (\beta -\mu)}\frac{\sin\pi \gamma}{\sin\pi \mu}
\hskip 0.2 cm  , \hskip 0.2 cm  B_2 ([m]_n)=e^{-i\pi \beta }\sqrt{1-
\left ( \frac{\sin\pi \gamma}{\sin\pi \mu} \right)^2}
\end{eqnarray*}
Therefore the relations between matrix elements in the l. h. s. and the r. h. s. 
 
of (\ref{3.3b}) are satisfied provided the constants $C_{[m]_n ,j}$ fulfill the
relations~:
\begin{equation}
\frac{\tilde{C}}{C^+}=\frac{C_{[m]_n +\Delta_n (j), l}\times C_{[m]_n , j}}
{C_{[m]_n +\Delta_n (l), j}\times C_{[m]_n , l}}=
\label{3.10}
\end{equation}
\begin{eqnarray*}
\sqrt{\frac{ (\mu +\gamma )\sin\pi (\mu -\gamma )}
{ (\mu -\gamma )\sin\pi (\mu +\gamma )}}\frac{\Gamma (1-\mu )\Gamma (1+\mu
-\gamma)}{\Gamma (1+\mu )\Gamma (1-\mu -\gamma)}
\end{eqnarray*}
The equation (\ref{3.10}) does not determine completely the constants 
$C_{[m]_n , j}$. A solution is~:
\begin{equation}
C_{[m]_n , j}=\sqrt{\frac{\prod_{j^\prime}\Gamma (1+\frac{p_{j^\prime n}-p_{jn}}{k+n})}{\prod_{j^\prime}\Gamma (1+\frac{p_{j^\prime n}-p_{jn}-1}{k+n})}} 
\label{3.11}
\end{equation}
The final form of the level zero matrix elements of
$u_{ij} (z)$ is therefore ( modulo an overall normalization constant)~:
\begin{equation}
\langle (\tilde{m})_n^{\infty}|u_{ij} (z)| (m)_n^0 \rangle =\frac{C_{[m]_n , j}
\langle (\tilde{m})_n |\gamma_{ij}| (m)_n \rangle}{z^{-h (\tilde{m})+h+h (m)}}
\label{result}
\end{equation}
 
These matrix elements are well-defined only if $[\tilde{m}]_n$ and $[m]_n$
satisfy the integrability condition (\ref{int}). It is easy to see that~:
\begin{equation}
\lim_{k\rightarrow \infty} \langle (\tilde{m})_n^{\infty}|u_{ij} (z)|
 (m)_n^0 \rangle =\langle (\tilde{m})_n |\gamma_{ij}| (m)_n \rangle
\end{equation}
i. e. in this limit we get the matrix elements of the unit tensor operators. 
This  coincides with the treatment of the "classical limit" of chiral $WZW$ in
\cite{GW}\cite{MS}. 
 
 Let us consider now the second case for the index structure of the correlation
function, when $ j=l$~:
\begin{equation}
\tilde{\Psi}_2 (z_1 , z_2 ) \equiv \langle (\tilde{m})_n^{\infty}|u_{ij}
 (z_1 )
u_{kj} (z_2 )| (m)_n^0 \rangle
\end{equation}
The highest weights of $ (\tilde{m})_n^{\infty}$ and $ (m)_n^0$
 are connected by~:
\begin{eqnarray*}
[\tilde{m}]_n=[m]_n +2\Delta_n (j)
\end{eqnarray*}
The space of invariant couplings $Inv (V^{\ast}_{\tilde{m}}\otimes V
\otimes V\otimes V_m )$ is one dimensional for the given
$| (\tilde{m})\rangle$, $| (m)\rangle$ and we can write~:
\begin{equation}
\tilde{\Psi}_2 (z_1 , z_2)=\langle (\tilde{m})_n |\gamma_{ij}\gamma_{kj}|
 (m)_n \rangle \tilde{\psi} (z_1 , z_2)\equiv \tilde{T}\tilde{\psi} (z_1 , z_2)
\end{equation}
where $\gamma_{pq}$ are the unit tensor operators. To simplify the KZ equations
we can use the following identities~:
\begin{mathletters}
\label{3.12}
\begin{equation}
 (t_1 \otimes t_2 )\tilde{T}= (1-\frac{1}{n})\tilde{T} \label{3.12a}
\end{equation}
\begin{equation}
 (t_1 \otimes t_m )\tilde{T}= (\frac{1}{n}-1+\frac{1}{2} (C_2 ([\tilde{m}]_n)
-C_2 ([m]_n +\Delta_n (j))-C_2))\tilde{T}\label{3.12b}
\end{equation}
\begin{equation}
 (t_2 \otimes t_m )\tilde{T}=\frac{1}{2} (C_2 ([m]_n +\Delta_n (j))
-C_2 ([m]_n)-C_2)\tilde{T}\label{3.12c}
\end{equation}
\end{mathletters}
After the substitution $\tilde{\psi} (z_1 , z_2 )= (z_1 z_2 )^{\frac
{C_2 ([\tilde{m}]_n)+C_2 ([m]_n)-2C_2}{4 (k+n)}}\tilde{\chi} (z)$ we get the
following equation~:
\begin{equation}
 (k+n)\frac{d\tilde{\chi}}{dz}=\left (\frac{2C_2 ([m]_n +\Delta_n (j))
-C_2 ([\tilde{m}]_n)-C_2 ([m]_n)}{4z}+\frac{1-n}{n (1-z)}\right)\tilde{\chi}
\end{equation}
Using the explicit expression for $C_2 ([m]_n)$ we obtain the general solution
for $\tilde{\chi} (z)$~:
\begin{equation}
\tilde{\chi} (z)=Cz^{\frac{1-n}{2n (k+n)}} (1-z)^{\frac{n-1}{n (k+n)}}
\end{equation}
The constant $C$ can be defined from the behavior of the $\tilde{\chi} (z)$
near the point $z=0$~:
\begin{equation}
C=C_{[m]_n +\Delta_n (j), j}\times C_{[m]_n , j}
\end{equation}
By the analytic continuation $z\rightarrow \frac{1}{z}$ we get~:
\begin{equation}
\tilde{\Psi} (\frac{1}{z})=e^{i\pi \frac{n-1}{n (k+n)}}\tilde{\Psi} (z)
\end{equation}
It  precisely coincides with the $B-$matrix exchange relation (\ref{3.4}). 
 
This ends the construction. We found the explicit form of  the $u_{ij} (z)$
operators  in terms of the projected vertex operators \cite{MS}. It is
given by equation (\ref{pvo}) , where the values of the constants $C_{[m]_n ,j}$
are calculated above (\ref{result}).

\subsection{Combining the left and right theories}

The situation for the WZW model is completely analogous to the finite
dimensional example. The "physical" Hilbert space ${\cal{H}}_{phys}$ is
defined by the condition~:
\begin{equation}
\forall \hskip 0.3 cm  |\psi \rangle \in {\cal H}_{phys}\hskip 0.3 cm
 (\vec{p}_l -\vec{p}_r)|\psi \rangle = 0
\end{equation}
From this follows that ${\cal{H}}_{phys}$ is~:
\begin{equation}
{\cal H}_{phys}= \bigoplus_{\{ \lambda \}} {\cal H}_{\lambda}^l \otimes
{\cal H}_{\lambda}^r
\end{equation}
where $\{ \lambda \}$ is the set of all integrable representations of
$\hat{SU} (n)_k$, the left and right highest weight in any given term being
the same. The physical fields, which commute with $\vec{\phi}$ are~:
\begin{equation}
g_{ij} (z, \bar{z})=\sum_k u_{ik} (z)v_{kj} (\bar{z})
\end{equation}
It can be shown that in the "physical" Hilbert space~:
\begin{equation}
[g_{ij} (z_1 , \bar{z}_1 ), g_{kl} (z_2 , \bar{z}_2 )]=0
\end{equation}
if $\eta (arg\frac{z_1}{z_2})=\eta (arg\frac{\bar{z}_1}{\bar{z}_2})$.

\section{Conclusion}
In this work  we constructed the Hilbert space for the $WZW$ , $\hat{SU} (n)_k$
model in the chiral quantization scheme. The algebra of the chiral group
operators can be realized on the space of all integrable representations of
$\hat{SU} (n)_k$ . The matrix elements of the
 chiral group elements $u_{ij} (z)$ in the fundamental representation
were found explicitly. 
 
We established the
connection between these operators and the projected vertex operators which
appear in the algebraic treatment of the chiral theory \cite{MS} . The chiral
group elements are  linear combinations with fixed coefficients  of the
projected vertex operators. 
Therefore they are not  invariant under the "gauge transformations" described
in the algebraic treatment \cite{MS}. The  significance of these particular
linear combinations of vertex operators from the algebraic point of view is
still unclear. Our lack of understanding of this feature is closely connected
with the missing  explanation
of the exact action of the quantum group symmetry in the chiral $WZW$.

We gave the exact rules for combining of the left and right chiral theories ,
following from the "Gauss law" implied by our gauge choice. The rules
 reproduce the standard treatment of the $WZW$ model \cite{KZ}\cite{GW}. 
 
The generalization of our treatment to the other groups may be a nontrivial
task. For the "toy model"  a  possible
difficulty  is connected with the generalization of the Gelfand pattern
notation ,
 which was crucial in our treatment. For the
$WZW$ model  the exact form of the commutation relations of the chiral
group elements is not known \cite{Dom} for other groups. 
 
Another open problem is the lagrangian meaning of the solutions
obtained combining in
a non diagonal way the left and right representations \cite{GFK}. 
 
{\bf Acknowledgments}

The work of M. L. was supported by Minerva Grant no. 8282 and by the Center 
for Basic Interactions. We would like to thank A. Schwimmer  for many helpful discussions. M. L. is grateful to M. S. Marinov for 
enlightening remarks about the motion on the group manifold  and to L. Feher 
for clarifying notes about chiral decomposition in $WZW$ model.

{\bf Appendix 1}
 
In this Appendix we will prove that
the commutation relations of the unit tensor operators
$\left\langle
\begin{array}{cc}
i\\
j\end{array}\right\rangle$
 are the same as the commutation relations of the $u_{ji}$ operators. 
 
Let us first consider the $U (n)$ tensor operators. Their definition  coincides
 with the definition of the $SU (n)$ tensor operators, but in this case $m_{nn}$
in
 the Gelfand pattern may be different from zero. The  proof can be given by
 induction. For $U (2)$ we have only four unit tensor operators (next to each
operator we indicate its  action on a generic state $| (m)\rangle$)~:
\begin{eqnarray*}
1)\hskip 1.5 cm  \left\langle\begin{array}{ccc}
1\\
1 0\\
1\end{array}\right\rangle \equiv \left\langle\begin{array}{cc}
1\\
1\end{array}\right\rangle \hskip 0.5 cm  ;\hskip 0.5 cm
\left\langle\begin{array}{cc}
1\\
1\end{array}\right\rangle\left|\begin{array}{cc}
m_{12}&m_{22}\\\
m_{11}\end{array}\right\rangle=
\end{eqnarray*}
\begin{eqnarray*}
\sqrt{\frac{m_{11}-m_{22}+1}{m_{12}-m_{22}+1}}
\left|\begin{array}{cc}
m_{12}+1&m_{22}\\\
m_{11}+1\end{array}\right\rangle
\end{eqnarray*}
\begin{eqnarray*}
2)\hskip 1.5 cm  \left\langle\begin{array}{ccc}
1\\
1 0\\
0\end{array}\right\rangle \equiv \left\langle\begin{array}{cc}
1\\
2\end{array}\right\rangle \hskip 0.5 cm  ;\hskip 0.5 cm
\left\langle\begin{array}{cc}
1\\
2\end{array}\right\rangle\left|\begin{array}{cc}
m_{12}&m_{22}\\\
m_{11}\end{array}\right\rangle=
\end{eqnarray*}
\begin{eqnarray*}
-\sqrt{\frac{m_{12}-m_{11}+1}{m_{12}-m_{22}+1}}
\left|\begin{array}{cc}
m_{12}+1&m_{22}\\\
m_{11}\end{array}\right\rangle
\end{eqnarray*}
\begin{eqnarray*}
3)\hskip 1.5 cm  \left\langle\begin{array}{ccc}
0\\
1 0\\
1\end{array}\right\rangle \equiv \left\langle\begin{array}{cc}
2\\
1\end{array}\right\rangle \hskip 0.5 cm  ;\hskip 0.5 cm
\left\langle\begin{array}{cc}
2\\
1\end{array}\right\rangle\left|\begin{array}{cc}
m_{12}&m_{22}\\\
m_{11}\end{array}\right\rangle=
\end{eqnarray*}
\begin{eqnarray*}
\sqrt{\frac{m_{12}-m_{11}}{m_{12}-m_{22}+1}}
\left|\begin{array}{cc}
m_{12}&m_{22}+1\\\
m_{11}+1\end{array}\right\rangle
\end{eqnarray*}
\begin{eqnarray*}
4)\hskip 1.5 cm  \left\langle\begin{array}{ccc}
0\\
1 0\\
0\end{array}\right\rangle \equiv \left\langle\begin{array}{cc}
2\\
2\end{array}\right\rangle \hskip 0.5 cm  ;\hskip 0.5 cm
\left\langle\begin{array}{cc}
2\\
2\end{array}\right\rangle\left|\begin{array}{cc}
m_{12}&m_{22}\\\
m_{11}\end{array}\right\rangle=
\end{eqnarray*}
\begin{eqnarray*}
\sqrt{\frac{m_{11}-m_{22}}{m_{12}-m_{22}+1}}
\left|\begin{array}{cc}
m_{12}&m_{22}+1\\\
m_{11}\end{array}\right\rangle
\end{eqnarray*}
By direct calculation it is possible to obtain (for$\hskip 0.3 cm  i \neq k$)~:
\begin{equation}
\left\langle\begin{array}{cc}
i\\
j\end{array}\right\rangle\left\langle\begin{array}{cc}
k\\
l\end{array}\right\rangle=\left\langle\begin{array}{cc}
i\\
l\end{array}\right\rangle\left\langle\begin{array}{cc}
k\\
j\end{array}\right\rangle \frac{\hbar}{p_i-p_k}+\left\langle\begin{array}{cc}
k\\
l\end{array}\right\rangle\left\langle\begin{array}{cc}
i\\
j\end{array}\right\rangle\sqrt{1-\frac{\hbar^2}{ (p_i-p_k)^2}}
\end{equation}
where
\begin{equation}p_i\left|\begin{array}{cc}
[m]_n\\
 (m)_{n-1}\end{array}\right\rangle=\hbar (m_{in}+n-i)\left|\begin{array}{cc}
[m]_n\\
 (m)_{n-1}\end{array}\right\rangle \equiv \hbar p_{in}\left|\begin{array}{cc}
[m]_n\\
 (m)_{n-1}\end{array}\right\rangle
\end{equation}
Let us assume that the statement is true for the $U (n-1)$ unit tensor operators. 
The product of two unit tensor operators  acts on the generic state
$| (m)\rangle$ as it follows from (\ref{2.9})~:
\begin{equation}
\left\langle\begin{array}{cc}
i\\
j\end{array}\right\rangle_n
\left\langle\begin{array}{cc}
k\\
l\end{array}
\right\rangle_n
\left|\begin{array}{cc}
[m]_n\\
 (m)_{n-1}\end{array}\right\rangle=\sum_{rs}
\left[\begin{array}{cc}
i & [m]_n +\Delta_n (k)\\
r & [m_{n-1}] + \Delta_{n-1} (s)\end{array}\right]
\left[\begin{array}{cc}
k & [m]_n\\
s & [m]_{n-1}\end{array}\right]
\left\langle\begin{array}{cc}
r\\
t\end{array}\right\rangle_{n-1}
\left\langle\begin{array}{cc}
s\\
q\end{array}\right\rangle_{n-1} \times\label{2.15}
\end{equation}
\begin{eqnarray*}
\times \left|\begin{array}{cc}
 
[m]_n+\Delta_n (i)+\Delta_n (k)\\\
 (m)_{n-1}\end{array}\right\rangle
\end{eqnarray*}
where the subscript $n\hskip 0.2cm ,\hskip 0.2cm n-1$ designates that the
relevant unit tensor
operators correspond to $U (n)\hskip 0.2cm ,\hskip 0.2cm U (n-1)$. 
 Equation (\ref{2.13}) gives~:
 
\begin{equation}
\left[\begin{array}{cc}
k & [m]_n\\
s & [m]_{n-1}\end{array}\right]
=\sqrt{\frac{\prod_{s^\prime \neq s}^{n-1} (p_{kn}-
p_{s^\prime, n-1})\prod_{k^\prime \neq k}^n (p_{s, n-1}-p_{k^\prime n}+1)}
{\prod_{k^\prime \neq k}^n (p_{kn}-p_{k^\prime n})\prod_{s^\prime \neq s}
^{n-1} (p_{s, n-1}-p_{s^\prime, n-1}+1)}}
\end{equation}
\begin{equation}
\left[\begin{array}{cc}
i & [m]_n +\Delta (k)_n\\
r & [m]_{n-1}+\Delta (s)_{n-1}\end{array}\right]
=\sqrt{\frac{\prod_{r^\prime \neq r}^{n-1} (p_{in}-
p_{r^\prime, n-1}-\delta_{r^\prime s})\prod_{i^\prime \neq i}^n (p_{r, n-1}
-p_{i^\prime n}-\delta_{i^\prime k}+1)}{\prod_{i^\prime \neq i}^n (p_{in}
-p_{i^\prime n}-\delta_{i^\prime k})\prod_{r^\prime \neq r}^{n-1} (p_{r, n-1}
-p_{r^\prime, n-1}-\delta_{r^\prime s}+1)}}\end{equation}
where $p_{ij}=m_{ij}+j-i$. By straightforward calculation it can be shown,
that~:
\begin{equation}
\left[\begin{array}{cc}
i & [m]_n +\Delta (k)_n\\
r & [m]_{n-1}+\Delta (s)_{n-1}\end{array}\right]
\left[\begin{array}{cc}
k & [m]_n\\
s & [m]_{n-1}\end{array}\right]=
C_1\left[\begin{array}{cc}
i & [m]_n +\Delta (k)_n\\
s & [m]_{n-1}+\Delta (r)_{n-1}\end{array}\right]
\left[\begin{array}{cc}
k & [m]_n\\
r & [m]_{n-1}\end{array}\right]+
\end{equation}
\begin{eqnarray*}
C_2\left[\begin{array}{cc}
k & [m]_n +\Delta (i)_n\\
s & [m]_{n-1}+\Delta (r)_{n-1}\end{array}\right]
\left[\begin{array}{cc}
i & [m]_n\\
r & [m]_{n-1}\end{array}\right]
\end{eqnarray*}
where
\begin{mathletters}
\label{2.16}
\begin{equation}
C_1=\frac{p_{in}-p_{kn}+p_{r, n-1}-p_{s, n-1}}{ (p_{in}-p_{kn})
\sqrt{ (p_{r, n-1}-p_{s, n-1}+1) (p_{r, n-1}-p_{s, n-1}-1)}}
\end{equation}
\begin{equation}
C_2=\frac{p_{r, n-1}-p_{s, n-1}}{p_{in}-p_{kn}}\sqrt{\frac{ (p_{in}-p_{kn}+1)
 (p_{in}-p_{kn}-1)}{ (p_{r, n-1}-p_{s, n-1}+1) (p_{r, n-1}-p_{s, n-1}-1)}}
\end{equation}
\end{mathletters}
if $r \neq s$ and
\begin{mathletters}
\label{2.17}
\begin{equation}
C_1=\frac{1}{p_{in}-p_{kn}}
\end{equation}
\begin{equation}
C_2=\frac{\sqrt{ (p_{in}-p_{kn}+1) (p_{in}-p_{kn}-1)}}{p_{in}-p_{kn}}
\end{equation}
\end{mathletters}
if $r=s$. By our assumption about $U (n-1)$ unit tensor operators~:
\begin{equation}
\left\langle\begin{array}{cc}
r\\
t\end{array}\right\rangle_{n-1}\left\langle\begin{array}{cc}
s\\
q\end{array}\right\rangle_{n-1}=\left\langle\begin{array}{cc}
r\\
q\end{array}\right\rangle_{n-1}\left\langle\begin{array}{cc}
s\\
t\end{array}\right\rangle_{n-1} \frac{1}{p_{r, n-1}-p_{s, n-1}}+
\end{equation}
\begin{eqnarray*}
\left\langle\begin{array}{cc}
s\\
q\end{array}\right\rangle_{n-1}\left\langle\begin{array}{cc}
r\\
t\end{array}\right\rangle_{n-1} \sqrt{1-\frac{1}{ (p_{r, n-1}-p_{s, n-1})^2}}
\end{eqnarray*}
if $r \neq s$ and
\begin{equation}
\left\langle\begin{array}{cc}
r\\
t\end{array}\right\rangle_{n-1}\left\langle\begin{array}{cc}
r\\
q\end{array}\right\rangle_{n-1}=\left\langle\begin{array}{cc}
r\\
q\end{array}\right\rangle_{n-1}
\left\langle\begin{array}{cc}
r\\
t\end{array}\right\rangle_{n-1} \label{2.18}
\end{equation}
By substitution of (\ref{2.16}), (\ref{2.18}) into (\ref{2.15})
 we obtain the desired result. The
 proof in the case, when $i=k$ or $j (l)=n$ is also straightforward and
 completely analogous to the previous one. For the $SU (n)$ unit tensor
 operators we will obtain  formally  the same result, if we will introduce the
 operator $p_n $; $p_n \equiv 0$ , because the only difference between an $U (n)$
 irrep $[m]_n$ and the corresponding $SU (n)$ irrep $[m]_n$ is the fact that
$m_{nn}=0$ for the  $SU (n)$ irrep. 
 
{\bf Appendix 2}
 
 Proof of relations (\ref{3.7}) , (\ref{3.12})
 
Let $t^a_{pq}$ denote the matrix elements of the $SU (n)$ generators in the
fundamental representation with normalization chosen as in the section
  \ref{s22} 
 and $t^a_{ (m) (m^\prime)}$ in the irrep $[m]_n$. Let $\langle (\tilde{m})_n |
\gamma_{ij} \gamma_{kl}| (m)_n \rangle$ denote the matrix elements of the
product of two unit tensor operators. Then~:
\begin{equation}
\sum_{a}\sum_{pq}t^a_{ip}t^a_{kq}\langle (\tilde{m})|\gamma_{pj}
\gamma_{ql}| (m)\rangle =
\end{equation}
\begin{eqnarray*}
\langle (\tilde{m})|\gamma_{kj}\gamma_{il}| (m)
\rangle -\frac{1}{n}\langle (\tilde{m})|\gamma_{ij}\gamma_{kl}| (m)\rangle
\end{eqnarray*}
which proves (\ref{3.7a}). Notice also that from (\ref{2.6})
 it  follows that~:
\begin{equation}
\langle (\tilde{m})|\gamma_{kj}\gamma_{ij}| (m)\rangle =\langle
 (\tilde{m})|\gamma_{ij}\gamma_{kj}| (m)\rangle 
\label{A1.1}
\end{equation}
which proves (\ref{3.12a}). We will prove now (\ref{3.7b}), (\ref{3.7c}). 
\begin{equation}
\sum_a \sum_{p, (m)^\prime }t^a_{ip} t^a_{ (m), (m^\prime )}\langle
 (\tilde{m}) |\gamma_{pj} \gamma_{kl}| (m^\prime )\rangle =
\end{equation}
\begin{eqnarray*}
\sum_a \langle
 (\tilde{m}) |[J^a , \gamma_{ij}]\gamma_{kl}J^a | (m)\rangle =
\end{eqnarray*}
 
\begin{eqnarray*}
\sum_a (-\langle (\tilde{m}) |[J^a , \gamma_{ij}][J^a , \gamma_{kl}]| (m)
\rangle +\langle (\tilde{m}) |[J^a , \gamma_{ij}]J^a \gamma_{kl}| (m)\rangle )=
\end{eqnarray*}

\begin{eqnarray*}
-\sum_a \sum_{pq} t^a_{ip}t^a_{kq}\langle (\tilde{m})|\gamma_{pj}
\gamma_{ql}| (m)\rangle +\frac{1}{2} (\langle (\tilde{m})|[\sum_a J^a J^a ,
\gamma_{ij}]\gamma_{kl}| (m)\rangle -
\end{eqnarray*}

\begin{eqnarray*}
\langle (\tilde{m})|\sum_a[J^a , [J^a \gamma_{ij}]]\gamma_{kl}|
 (m)\rangle )=
-\langle (\tilde{m})| \gamma_{kj}\gamma_{il}| (m)\rangle +
\end{eqnarray*}
\begin{eqnarray*}
\frac{1}{2} (\frac{2}{n}+C_2 ([\tilde{m}]_n)-C_2 ([m]_n +\Delta_n (l))-C_2)
\langle (\tilde{m})|\gamma_{ij}\gamma_{kl}| (m)\rangle
\end{eqnarray*}

To show (\ref{3.12b}) we must take into account (\ref{A1.1}). 
\begin{equation}
\sum_a \sum_{q, (m)^\prime }t^a_{kq} t^a_{ (m), (m^\prime )}\langle
 (\tilde{m}) |\gamma_{ij} \gamma_{ql}| (m^\prime )\rangle =
\end{equation}
\begin{eqnarray*}
\sum_a \langle (\tilde{m}) |\gamma_{ij}[J^a , \gamma_{kl}]J^a | (m)\rangle =
\end{eqnarray*}
\begin{eqnarray*}
\frac{1}{2} (\langle (\tilde{m})|\gamma_{ij}[\sum_a J^a J^a , \gamma_{kl}]| (m)
\rangle -\langle (\tilde{m})|\gamma_{ij}\sum_a[J^a , [J^a \gamma_{kl}]]| (m)
\rangle )=
\end{eqnarray*}
\begin{eqnarray*}
\frac{1}{2} (C_2 ([m]_n +\Delta_n (l))-C_2 ([m]_n)-C_2)
\langle (\tilde{m})|\gamma_{ij}\gamma_{kl}| (m)\rangle
\end{eqnarray*}
 
which proves (\ref{3.7d}), (\ref{3.7e}). 
 
 {\bf Appendix 3}

 Proof of the formulas (\ref{3.9}).

We use the following identities of the hypergeometric
 functions\cite{Sp}~:
\begin{mathletters}
\label{hyperid}
\begin{equation}
 (\gamma -\alpha -1)F (\alpha , \beta , \gamma , z)+\alpha F (\alpha +1,
\beta ,
\gamma , z)-
\end{equation}
\begin{eqnarray*}
 (\gamma -1)F (\alpha , \beta , \gamma -1, z)=0
\end{eqnarray*}
\begin{equation}
 (\gamma -\beta -1)F (\alpha , \beta , \gamma , z)+\beta F (\alpha ,
\beta +1,
\gamma , z)-
\end{equation}
\begin{eqnarray*}
 (\gamma -1)F (\alpha , \beta , \gamma -1, z)=0
\end{eqnarray*}
\end{mathletters}
From (\ref{hyperid}) it follows that~:
\begin{equation}
F (\alpha , \beta , \gamma , z)=\frac{\alpha (\gamma -\beta )}{ (\beta -1)
 (\gamma -\alpha -1)} F (\alpha +1 , \beta -1, \gamma , z)+
\end{equation}
\begin{eqnarray*}
\frac{ (\gamma -1) (\beta -\alpha )}{ (\beta -1) (\gamma -\alpha -1)}
F (\alpha , \beta -1, \gamma -1, z)
\end{eqnarray*}
and by substitution of the explicit expression for $\xi_1 (z)$ and
$\chi^-_{1, 2} (z)$ we will obtain (\ref{3.9a}). 
 From the~:
\begin{mathletters}
\begin{equation}
 (\gamma -\alpha -1)F (\alpha , \beta , \gamma , z)+
\end{equation}
\begin{eqnarray*}
\alpha F (\alpha +1, \beta ,
\gamma , z)- (\gamma -1)F (\alpha , \beta , \gamma -1, z)=0
\end{eqnarray*}
\begin{equation}
\beta zF (\alpha , \beta +1, \gamma +1, z)=\gamma F (\alpha , \beta ,
\gamma , z)-\gamma F (\alpha -1, \beta , \gamma , z)
\end{equation}
\end{mathletters}
it follows that~:
\begin{equation}
zF (\alpha , \beta +1, \gamma +1, z)=
\end{equation}
\begin{eqnarray*}
\frac{\gamma (1-\gamma)}{\beta (\alpha -
\gamma)} ( F (\alpha , \beta , \gamma , z)-F (\alpha -1, \beta , \gamma -1, z))
\end{eqnarray*}
which proves (\ref{3.9b}).

\end{document}